\documentclass[conference,compsoc]{IEEEtran}

\usepackage{algorithm}
\usepackage{algorithmic}
\usepackage{amsmath,amsfonts,amssymb}
\usepackage{hyperref}
\usepackage{graphicx}
\usepackage{sidecap}
\usepackage{caption}
\usepackage{subcaption}
\usepackage{verbatim}

\sidecaptionvpos{figure}{t}

\newcommand{\mtx}[1]{\ensuremath{\mathbf{#1}}}

\DeclareMathOperator*{\argmin}{arg\,min} 

\ifCLASSOPTIONcompsoc
  \usepackage[nocompress]{cite}
\else
  \usepackage{cite}
\fi

\ifCLASSINFOpdf
\else
\fi

\begin{document}

\title{Hybrid Collaborative Filtering with Autoencoders}


\author{\IEEEauthorblockN{Florian Strub}
\IEEEauthorblockA{Univ. Lille, CNRS, Centrale Lille\\ 
UMR 9189 - CRIStAL\\
F-59000 Lille, France \\
Email: florian.strub@inria.fr}
\and
\IEEEauthorblockN{J\'er\'emie Mary}
\IEEEauthorblockA{Univ. Lille, CNRS, Centrale Lille\\ 
UMR 9189 - CRIStAL\\
F-59000 Lille, France \\
Email: jeremie.mary@inria.fr}
\and
\IEEEauthorblockN{Romaric Gaudel}
\IEEEauthorblockA{Univ. Lille, CNRS, Centrale Lille\\ 
UMR 9189 - CRIStAL\\
F-59000 Lille, France \\
Email: romaric.gaudel@inria.fr}
}

\maketitle

\begin{abstract} Collaborative Filtering aims at exploiting the feedback of
users to provide personalised recommendations. Such algorithms look for latent
variables in a large sparse matrix of ratings. They can be enhanced by adding
side information to tackle the well-known cold start problem. While Neural
Networks have tremendous success in image and speech recognition, they have
received less attention in Collaborative Filtering. This is all the more
surprising that Neural Networks are able to discover latent variables in large
and heterogeneous datasets. In this paper, we introduce a Collaborative
Filtering Neural network architecture aka CFN which computes a non-linear
Matrix Factorization from sparse rating inputs and side information. We show
experimentally on the MovieLens and Douban dataset that CFN outperforms the
state of the art and benefits from side information. We provide an
implementation of the algorithm as a reusable plugin for Torch, a popular
Neural Network framework. \end{abstract}

\IEEEpeerreviewmaketitle

\section{Introduction}
Recommendation systems advise users on which items (movies, musics, books etc.) they are more likely to be interested in.
A good recommendation system may dramatically increase the amount of sales of a firm or retain customers.
For instance, 80\% of movies watched on Netflix come from the recommender system of the company \cite{Netflix2015}.
One efficient way to design such algorithm is to predict how a user would rate a given item. 
Two key methods co-exist to tackle this issue: \emph{Content-Based Filtering} (CBF) and \emph{Collaborative Filtering} (CF).

CBF uses the user/item knowledge to estimate a
new rating. For instance, user information can be the age, gender, or
graph of friends etc. Item information can be the movie genre, a short
description, or the tags. On the other side, CF uses the ratings history 
of users and items. The feedback of \emph{one} user on \emph{some} items is
combined with the feedback of \emph{all} other users on \emph{all} items to
predict a new rating. For instance, if someone rated a few books, Collaborative
Filtering aims at estimating the ratings he would have given to thousands of
other books by using the ratings of all the other readers. CF
is often preferred to CBF because it wins the
agnostic vs. studied contest: CF only relies on the
ratings of the users while CBF requires advanced
engineering on items to well perform \cite{Lops2011}.

The most successful approach in CF is to retrieve
potential latent factors from the sparse matrix of ratings. Book latent factors
are likely to encapsulate the book genre (spy novel, fantasy, etc.) or some
writing styles. Common latent factor techniques compute a low-rank
rating matrix by applying Singular Value Decomposition through gradient descent
\cite{Koren2009} or Regularized Alternating Least Square algorithm
\cite{Zhou2008}. However, these methods are linear and
cannot catch subtle factors. Newer algorithms were explored to face those
constraints such as Factorization Machines \cite{Rendle2010}. 
More recent works combine several low-rank matrices 
such as Local Low Rank Matrix Approximation \cite{Lee2013} or WEMAREC \cite{Chen2015} to enhance the recommendation.

Another limitation of CF is known as the \emph{cold start} problem: how to
recommend an item to a user when no rating exists for neither the user nor the item?
To overcome this issue, one idea
is to build a hybrid model mixing CF and CBF
where side information is integrated into the training process. The
goal is to supplant the lack of ratings through side information. A
successful approach \cite{Adams2010,Porteous2010} extends the Bayesian
Probabilistic Matrix Factorization Framework \cite{Salakhutdinov2008} to
integrate side information. However, recent algorithms outperform them in the general case \cite{Lee2012}.

In this paper we introduce a CF approach based on Stacked Denoising Autoencoders
\cite{Vincent2010} which tackles
both challenges: learning a non-linear representation of users and items, and
alleviating the cold start problem by integrating side information. Compared to
previous attempts in that direction
\cite{Salakhutdinov2007,Sedhain2015,Strub2015,Dziugaite2015,Wu2016},
our framework integrates the sparse matrix of ratings and side information in a unique Network.
This joint model leads to a scalable and robust approach which beats state-of-the-art results in CF.
Reusable source code is provided in Torch to reproduce the results.
Last but not least, we show that CF approaches based on Matrix Factorization
have a strong link with our approach.

The paper is organized as follows. First, Sec.\@ \ref{sec:preliminaries} summarizes the
state-of-the-art in CF and Neural Networks.
Then, our model is described in Sec.\@ \ref{sec:model} and \ref{sec:side}
and its relation with
Matrix Factorization is characterized in Sec.\@ \ref{sec:NNvsMF}.
Finally, experimental results are given and discussed in Sec.\@ \ref{sec:exp}
and Sec.\@ \ref{sec:remarks} discusses algorithmic aspects.

\section{Preliminaries}
\label{sec:preliminaries}

\subsection{Denoising Autoencoders}
The proposed approach builds upon Autoencoders which are feed-forward Neural Networks popularized by Kramer \cite{Kramer1991}. They are unsupervised Networks where the output of the Network aims at reconstructing the initial input. The Network is constrained to use narrow hidden layers, forcing a dimensionality reduction on the data. The Network is trained by back-propagating the squared error loss on the reconstruction. Such Networks are divided into two parts:
\begin{itemize} 
\item the encoder : $f(\mtx{x}) = \sigma(\mtx{W_1 x + b_1})$,
\item the decoder : $g(\mtx{y}) = \sigma(\mtx{W_2 y + b_2})$,
\end{itemize}
with $\mtx{x} \in \mathbb{R}^{N}$ the input, $\mtx{y} \in \mathbb{R}^{k}$ the
output, $k$ the size of the Autoencoder's bottleneck ($k \ll N$), $\mtx{W_1} \in
\mathbb{R}^{k\times N}$ and $\mtx{W_2} \in \mathbb{R}^{N\times k}$ the weight
matrices, $\mtx{b_1} \in \mathbb{R}^{k}$ and $\mtx{b_2} \in \mathbb{R}^{N}$ the
bias vectors, and $\sigma(.)$ a non-linear transfer function. The full Autoencoder will be denoted $nn(\mtx{x}) \stackrel{\text{def}}{=} g(f(\mtx{x}))$.

Recent work in Deep Learning advocates to stack pre-trained encoders to
initialize Deep Neural Networks \cite{Glorot2010}. This process enables the
lowest layers of the Network to find low-dimensional representations. It
experimentally increases the quality of the whole Network. Yet, classic
Autoencoders may degenerate into identity Networks and they fail to learn the
latent relationship between data. \cite{Vincent2010} tackle this issue by corrupting
inputs, pushing the Network to denoise the final outputs. One method is to add
Gaussian noise on a random fraction of the input. Another method is to
mask a random fraction of the input by replacing them with zero.
In this case, the Denoising AutoEncoder (DAE) loss function is modified to
emphasize the denoising aspect of the Network. The loss is based on two main
hyperparameters $\alpha$, $\beta$. They balance whether the Network would focus
on denoising the input ($\alpha$) or reconstructing the input ($\beta$):
\begin{multline*} L_{2,\alpha, \beta}(\mtx{x},\mtx{\tilde{x}}) =
\alpha\left(\sum_{j\in \mathcal{C}(\mtx{\tilde{x})}}[nn(\mtx{\tilde{x}})_j -
x_j]^2\right) + \\ \beta
\left(\sum_{j\not\in\mathcal{C}(\mtx{\tilde{x})}}[nn(\mtx{\tilde{x}})_j -
x_j]^2\right), \end{multline*} 
where $\mtx{\tilde{x}} \in \mathbb{R}^{N}$ is a corrupted version of the input
$\mtx{x}$, $\mathcal{C}$ is the set of corrupted
elements in $\mtx{\tilde{x}}$, and $nn(\mtx{x})_j$ is the $j^{th}$ output of the Network while fed with $\mtx{x}$.

\subsection{Matrix Factorization}
One of the most successful approach of Collaborative Filtering is Matrix Factorization \cite{Koren2009}. This method retrieves latent factors from the ratings given by the users. The underlying idea is that key features are hidden in the ratings themselves. 
Given $N$ users and $M$ items, the rating $r_{ij}$ is the rating given by the $i^{th}$  user for the $j^{th}$ item. 
It entails a sparse matrix of ratings $\mtx{R} \in \mathbb{R}^{N\times M}$. In Collaborative Filtering, sparsity is originally produced by missing values
rather than zero values. The goal of Matrix Factorization is to find a $k$ low rank matrix $\mtx{\widehat{R}} \in \mathbb{R}^{N\times M}$ where  $\mtx{\widehat{R}} = \mtx{U}\mtx{V}^{T}$ with $\mtx{U} \in \mathbb{R}^{N\times k}$ and $\mtx{V} \in \mathbb{R}^{M\times k}$ two matrices of rank $k$ encoding a dense representation of the users/items. In it simplest form, ($\mtx{U}$ ,$\mtx{V}$) is the solution of
  $$\argmin_{\mtx{U},\mtx{V}}
  \sum_{(i,j) \in \mathcal{K}(\mtx{R}) }  (r_{ij} - \mtx{\bar{u}}_i^T\mtx{\bar{v}}_j)^2 + \lambda( \|\bar{\mtx{u}}_i \|_{F}^2 + \|\bar{\mtx{v}}_j \|_{F}^2),$$
where 
 $\mathcal{K}(\mtx{R})$ is the set of indices of known ratings of $\mtx{R}$,
 ($\bar{\mtx{u}}_i$, $\bar{\mtx{v}}_j$) are the dense and low rank rows of ($\mtx{U}$ ,$\mtx{V}$) and 
 $\|.\|_{F}$ is the Frobenius norm. Vectors $\bar{\mtx{u}}_i$ and $\bar{\mtx{v}}_j$ are treated as column-vectors.

\begin{figure*}[t]
\begin{center}
\centerline{\includegraphics[width=\linewidth]{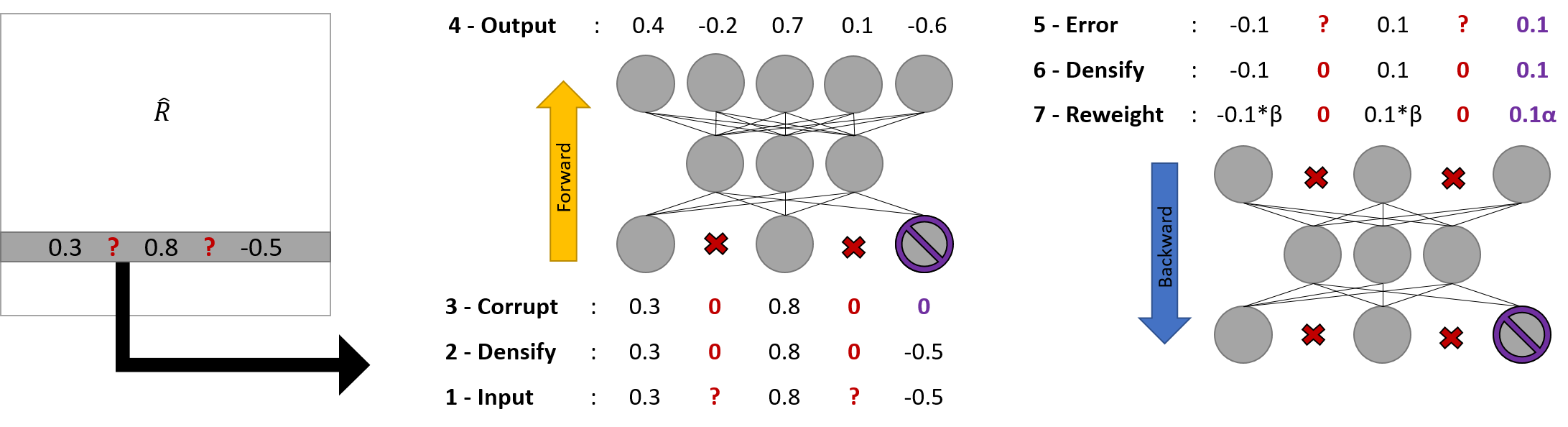}}
\caption{Feed Forward/Backward process for sparse Autoencoders. The sparse input is drawn from the matrix of ratings, unknown values are turned to zero, some ratings are masked (input corruption) and a dense estimate is finally obtained. Before backpropagation, unknown ratings are turned to zero error, prediction errors are reweighed by $\alpha$ and reconstruction errors are reweighed by $\beta$.}
\label{img:training}
\vskip -0.5em
\end{center}
\end{figure*}

\subsection{Related Work} 
Neural Networks have attracted little attention in the CF
community. In a preliminary work, \cite{Salakhutdinov2007} tackled the Netflix
challenge using Restricted Boltzmann Machines but little published work had follow \cite{Phung2009}.
While Deep Learning has tremendous success in image and speech recognition \cite{Lecun2015},
sparse data has received less attention and remains a challenging problem for
Neural Networks.

Nevertheless, Neural
Networks are able to discover non-linear latent variables with heterogeneous
data \cite{Lecun2015} which makes them a promising tool for CF.
\cite{Sedhain2015,Strub2015,Dziugaite2015} directly train Autoencoders to
provide the best predicted ratings. Those methods
report excellent results in the general case. However, the cold start
initialization problem is ignored. For instance, AutoRec \cite{Sedhain2015}
replaces unpredictable ratings by an arbitrary selected score.
In our case, we apply a training loss designed for \emph{sparse} rating inputs
and we integrate side information to lessen the cold start effect.

Other contributions
deal with this cold start problem by using Neural Networks properties for
CBF: Neural Networks are first trained to learn a feature
representation from the item which is then processed by a CF approach such as
Probabilistic Matrix Factorization \cite{Mnih2007} to provide the
final rating. For instance, \cite{Glorot2011,Wang2014} respectively auto-encode bag-of-words from
restaurant reviews and movie plots, \cite{Li2015} auto-encode heterogeneous side
information from users and items. Finally, \cite{Van2013,Wang2014b} use Convolutional Networks on music
samples. In our case, side information and ratings are used together without any
unsupervised pretreatment.

\subsection{Notation}
In the rest of the paper, we will use the following notations:

\begin{itemize} 
\item $\mtx{u_i}$, $\mtx{v_j}$ are the sparse rows/columns of $\mtx{R}$;
\item $\mtx{\tilde{u}}_i$, $\mtx{\tilde{v}_j}$ are corrupted versions of $\mtx{u_i}$, $\mtx{v_j}$;
\item $\mtx{\hat{u}}_i$, $\mtx{\hat{v}}_j$ are dense estimates of $\mtx{\widehat{R}}$;
\item $\mtx{\bar{u}}_i$, $\mtx{\bar{v}}_j$ are dense low rank representations of $\mtx{u_i}$, $\mtx{v_j}$.
\end{itemize}

\section{Autoencoders and CF}
\label{sec:model}

User preferences are encoded as a sparse matrix of ratings $\mtx{R}$. A user is represented by a sparse line $\mtx{u}_i \in \mathbb{R}^{N}$ and an item is represented by a sparse column $\mtx{v}_j \in \mathbb{R}^{M}$. The Collaborative Filtering objective can be formulated as: \emph{turn the sparse vectors $\mtx{u}_i$/$\mtx{v}_j$, into dense vectors $\hat{\mtx{u}}_i$/$\hat{\mtx{v}}_j$}.

We propose to perform this conversion with Autoencoders. To do so, we need to define two types of Autoencoders:

\begin{itemize} 
\item U-CFN is defined as $\hat{\mtx{u}}_i = nn(\mtx{u_i})$,
\item V-CFN is defined as $\hat{\mtx{v}}_j = nn(\mtx{v_j})$.  
\end{itemize}

The encoding part of these Autoencoders aims at building a low-rank dense
representation of the sparse input of ratings. The decoding part aims at
predicting a dense vector of ratings from the low-rank dense representation of
the encoder. This new approach differs from classic Autoencoders which only
aim at reconstructing/denoising the input. As we will see later, the
training loss will then differ from the evaluation one.

\subsection{Sparse Inputs}
\label{sec:sparse}
There is no standard approach for using sparse vectors as inputs of Neural
Networks. Most of the papers dealing with sparse inputs get around by
pre-computing an estimate of the missing values \cite{Tresp1994,Bishop1995}.
In our case, we want the Autoencoder to handle this prediction issue by itself.
Such problems have already been studied in industry \cite{Miranda2012} where
5\% of the values are missing. However in Collaborative Filtering we often face
datasets with more than 95\% missing values. Furthermore, missing values are
\emph{not} known during \emph{training} in Collaborative Filtering which
makes the task even more difficult.  

Our approach  includes three ingredients to handle the training of sparse Autoencoders:
\begin{itemize} 
\item inhibit the edges of the input layers by zeroing out values in the input,
\item inhibit the edges of the output layers by zeroing out back-propagated values,
\item use a denoising loss to emphasize rating prediction over rating reconstruction. 
\end{itemize}

One way to inhibit the input edges is to turn missing values to zero. To keep
the Autoencoder from always returning zero, we also use an empirical loss that
disregards the loss of unknown values. No error is back-propagated for missing
values. Therefore, the error is back-propagated for actual
zero values while it is discarded for missing values. In other words, missing
values do not bring information to the Network. This operation is equivalent to
removing the neurons with missing values described in \cite{Salakhutdinov2007,Sedhain2015}. 
However, Our method has important computational advantages because only
one Neural Networks is trained whereas other techniques has to share the weights among thousands of
Networks.

Finally, we take advantage of the masking noise from the Denoising AutoEncoders (DAE) empirical
loss. By simulating missing values in the training process, Autoencoders are
trained to predict them. In Collaborative Filtering, this prediction aspect is
actually the final target. Thus, emphasizing the prediction criterion turns the
classic unsupervised training of Autoencoders into a simulated supervfigureised
learning. By mixing both the reconstruction and prediction criteria, the
training can be thought as a pseudo-semi-supervised learning. This makes the
DAE loss a promising objective function.
After regularization, the final training loss is:
\begin{multline*}
L_{2,\alpha, \beta}(\mtx{x},\mtx{\tilde{x}}) = 
\alpha\left(\sum_{j\in    \mathcal{C}(\tilde{\mtx{x}})\cap\mathcal{K}(\mtx{x})}[nn(\mtx{\tilde{x}})_j - x_j]^2\right)
 + \\
\beta \left(\sum_{j\not\in\mathcal{C}(\tilde{\mtx{x}})\cap\mathcal{K}(\mtx{x})}[nn(\mtx{\tilde{x}})_j - x_j]^2\right) + \lambda\|\mtx{W}\|_{F}^2,
\end{multline*}
%
where $\mathcal{K}(\mtx{x})$ are the indices of known values of $\mtx{x}$, $\mtx{W}$ is the flatten vector of weights of the Network and $\lambda$ is the regularization hyperparameter. The full forward/backward process is explained in Figure \ref{img:training}. Importantly, Autoencoders with sparse inputs differs from sparse-Autoencoders \cite{Lee2006} or Dropout regularization \cite{Srivastava2014} in the sense that Sparse Autoencoders and Droupout inhibit the hidden neurons for regularization purpose. Every inputs/outputs are also known.

\subsection{Low Rank Matrix Factorization}
\label{sec:NNvsMF}

Autoencoders are actually strongly linked with Matrix Factorization. For an Autoencoder with only one hidden layer and no output transfer function, the response of the network is
$
nn(\mtx{x}) = \mtx{W_2} \; \sigma(\mtx{W_1x + b_1}) + \mtx{b_2},
$
where $\mtx{W_1,W_2}$ are the weights matrices and $\mtx{b_1,b_2}$ the bias
terms. Let $\mtx{x}$ be the representation
$\mtx{u}_i$ of the user $i$, then we recover a predicted vector $\hat{\mtx{u}}_i$
of the form $\mtx{V}\mtx{\bar{u}_i}$:
\begin{equation*}
\hat{\mtx{u}}_i
= nn(\mtx{u_i})
= \underbrace{\left[\mtx{W_2} \; {\mtx{I}_N}\right]}_{\mtx{V}}\;
\underbrace{\left[\begin{array}{c}\sigma(\mtx{W_1u_i + b_1})\\\mtx{b_2}\end{array}
\right]}_{\mtx{\bar{u}_i}}.
\end{equation*}
Symmetrically, $\hat{\mtx{v}}_j$ has the form $\mtx{U}\mtx{\bar{v}_j}$:
$$
\hat{\mtx{v}}_j =
nn(\mtx{v_j}) =
\underbrace{\left[\mtx{W'_2}\;\mtx{I}_M\right]}_{\mtx{U}} \;
\underbrace{\left[\begin{array}{c}\sigma(\mtx{W'_1v_j + b'_1})\\\mtx{b'_2}\end{array}\right]}_{\mtx{\bar{v}_j}}.
$$

The difference with standard Low Rank Matrix Factorization stands in the
definition of $\mtx{\bar{u}_i}$/$\mtx{\bar{v}_j}$.
For the Matrix Factorization by ALS, $\mtx{\widehat{R}}$ is iteratively built by solving for each row $i$ of $\mtx{U}$ (resp. column $j$ of $\mtx{V^T}$) a {\bf linear} least square regression using the known values of the $i^{th}$ row of  $\mtx{R}$ (resp. $j^{th}$ column of $\mtx{R}$) as observations of a scalar product in dimension $k$ of $\mtx{\bar{u}_i}$ and the corresponding columns of $\mtx{V^T}$ (resp. $\mtx{\bar{v}_j}$ and the corresponding rows of $\mtx{U}$). An Autoencoder aims at a projection in dimension $k$ composed with the non linearity $\sigma$. This process corresponds to a {\bf non linear} matrix factorization.

Note that CFN also breaks the symmetry between $\mtx{U}$ and $\mtx{V}$.
For example, while Matrix Factorization approaches learn both $\mtx{U}$ and $\mtx{V}$,
U-CFN learns $\mtx{V}$ and only indirectly learns $\mtx{U}$: U-CFN targets the
function to build $\mtx{\bar{u}_i}$ whatever the row $\mtx{u_i}$.
A nice benefit is that the learned Autoencoder is able to fill in every vector
$\mtx{u_i}$, even if that vector was not in the training data.

Both non-linear decompositions on rows and columns are done independently,
which means that the matrix $\mtx{V}$ learned by U-CFN from rows can differ from the
concatenation of vectors $\mtx{\bar{v}_j}$ predicted by V-CFN from columns. 

Finally, it is very important to differentiate CFN from Restrictive Boltzman Machine (RBM) for Collaborative Filtering \cite{Salakhutdinov2007}. By construction, RBM only handles binary input. Thus, one has to discretize the rating of users/items for both the input/output layers. First, it striclty limits the use of RBM on database with real numbers. Secondly, the resulting weight architecture clearly differs from CFN. in RBM, Imput/output ratings are encoded by $D$ weights where $D$ is the number of discretized features while CFN only requires a single weight. Thus, no direct link can be done between Matrix Factorization and RBM . Besides, this architecture also prevents RBM from being used to initialize the input/ouput layers of CFN.

\section{Integrating side information}
\label{sec:side}


\begin{figure}[t]
\centering
\includegraphics[width=0.8\linewidth]{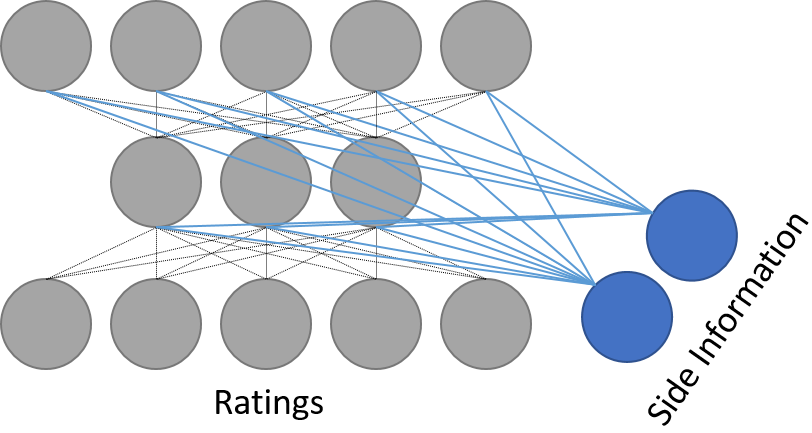}
\caption{Integrating side information. The Network has two inputs: the classic Autoencoder rating input and a side information input. Side information is wired to every neurons in the Network.}
\label{fig:side_info}
\end{figure}

 Collaborative Filtering only relies on the feedback of users regarding a set of items. When additional information
is available for the users and the items, this can sound restrictive. One would
think that adding more information can help in several ways: increasing the
prediction accuracy, speeding up the training, increasing the robustness of the
model, etc. Furthermore, pure Collaborative Filtering suffers from the cold start
problem: when very little information is available on an item, Collaborative Filtering will have difficulties recommending it.
When bad recommendations are provided, the probability to receive valuable feedback is lowered leading to a vicious circle for new items. A common way to tackle this problem is to add
some side information to ensure a better initialization of the system. This is known in the recommendation community as hybridization. 


The simplest approach to integrate side information is to append additional user/item bias to the rating prediction \cite{Koren2009}: 
$$\hat{r}_{ij} = \bar{\mtx{u}}_i^T\bar{\mtx{v}}_j + b_{u,i} + b_{v,j} + b',$$
where $b_{u,i}$, $b_{v,j}$, $b'$ are respectively the user, item, and global bias of the Matrix Factorization. Computing these bias can be done through hand-crafted engineering or Collaborative Filtering technique. For instance, one method is to extend the dense feature vectors of rank $k$ by directly appending side information on them \cite{Porteous2010}. Therefore, the estimated rating is computed by:
\begin{align*}
\hat{r}_{ij} &= \{\bar{\mtx{u}}_i, \mtx{x}_i\} \otimes \{\bar{\mtx{v}}_j, \mtx{y}_j\}\\
&\stackrel{\text{def}}{=} 
\bar{\mtx{u}}_{[1:k], i}^T\bar{\mtx{v}}_{[1:k], j} + 
\underbrace{\bar{\mtx{u}}_{[k+1:k+Q], i}^T\mtx{y}_j}_{b_{v,j}} + 
\underbrace{\mtx{x}_i^T\bar{\mtx{v}}_{[k+1:k+P], j}}_{b_{u,i}},
\end{align*}
where $\mtx{x}_i \in \mathbb{R}^{P}$ and $\mtx{y}_j \in \mathbb{R}^{Q}$ respectively are a vector representation of side information for the user and for the item. 
Unfortunately, those methods cannot be directly applied to Neural Networks 
because Autoencoders optimize $\mtx{U}$ and $\mtx{V}$ independently. 
New strategies must be designed to incorporate side information. 
One notable example  was recently made by \cite{Ammar2014} for bitext word alignment. 

In our case, the first idea would be to append the side information to the sparse input vector. For simplicity purpose, the next equations will only focus on shallow U-Autoencoders with no output transfer functions. Yet, this can be extended to more complex Networks and V-Autoencoders. Therefore, we get:
\begin{align*}
\hat{\mtx{u}}_i = nn(\{\mtx{u}_i, \mtx{x}_i\}) 
                             &=  \mtx{V} \;\sigma( \mtx{W}'_1 \{\mtx{u}_i, \mtx{x}_i\} + \mtx{b_1} )  + \mtx{b_2},
\end{align*}
where $\mtx{W}'_1 \in \mathbb{R}^{k \times (N+P)} $ is a weight matrix.

When no previous rating exist, it enables the Neural Networks to have at an input to predict new ratings. With this scenario, side information is assimilated to pseudo-ratings that will always exist for every items. However, when the dimension of the Neural Network input is far greater than the dimension of the side information, the Autoencoder may have difficulties to use it efficiently.

Yet, common Matrix Factorization would append side information to dense feature representations $\{\mtx{\bar{u}}_i, \mtx{x}_i\}$ rather than sparse feature representation as we just proposed $\{\mtx{u}_i, \mtx{x}_i\}$. A solution to reproduce this idea is to inject the side information to every layer inputs of the Network:
\begin{align*}
nn(\{ \mtx{u}_i, \mtx{x}_i\})&= \mtx{V}' \;
\{\overbrace{\sigma( \mtx{W}'_1 \{\mtx{u}_i, \mtx{x}_i\} + \mtx{b_1} )}^{\mtx{\bar{u}'}_i}, \mtx{x}_i\}  + \mtx{b_2} \\
&= \mtx{V}' \; \{ \mtx{\bar{u}'}_i, \mtx{x}_i\} + \mtx{b_2} \\
&= \mtx{V}'_{[1:k]}\mtx{\bar{u}'}_i +  \underbrace{\mtx{V}'_{[k+1:k+P]}\mtx{x}_i}_{\mtx{b}_{u,i}} + \mtx{b_2},
\end{align*}
where $\mtx{V}' \in \mathbb{R}^ {(N \times k+P)}$ is a weight matrix, $\mtx{V}'_{[1:k]} \in \mathbb{R}^ {N \times k}, \mtx{V}'_{[k+1:k+P]} \in \mathbb{R}^ {N \times P}$ are respectively the submatrices of $\mtx{V}'$ that contain the columns from $1$ to $k$ and $k+1$ to $k+P$. 

By injecting the side information in every layer, the dynamic Autoencoders
representation is forced to integrate this new data. 
However, to avoid side information to overstep the dense rating representation. Thus, we
enforce the following constraint. The dimension of the sparse input must be
greater than the dimension of the Autoencoder bottleneck which must be greater
than the dimension of the side information \footnote{When side information is
sparse, the dimension of the side information can be assimilated to the number
of non-zero parameters}. Therefore, we get: 
\begin{align*} P \ll k \ll N \;\; \; \text{ and } \;\;
\; Q \ll k \ll M. \end{align*}

We finally obtain an Autoencoder which can incorporate side information and be trained through backpropagation. See Figure \ref{fig:side_info} for a graphical representation of the corresponding network.


\section{Experiments}
\label{sec:exp}

\subsection{Benchmark Models}
We benchmark CFN with five matrix completion algorithms:
\begin{itemize}
\item
ALS-WR (Alternating Least Squares with Weighted-$\lambda$-Regularization) \cite{Zhou2008} solves the low-rank matrix factorization problem by alternatively fixing $\mtx{U}$ and $\mtx{V}$ and solving the resulting linear regression problem. Experiments are run with the Apache Mahout\footnote{\url{http://mahout.apache.org/}}. We use a rank of 200;
 \item SVDFeature \cite{Chen2012} learns a feature-based matrix factorization: side information are used to predict the bias term and to reweight the matrix factorization. We use a rank of 64 and tune other hyperparameters by random search;
  \item BPMF (Bayesian Probabilistic Matrix Factorization) \cite{Salakhutdinov2008} infers the matrix decomposition after a statistical model. We use a rank of 10;
 \item LLORMA \cite{Lee2013} estimates the rating matrix as a weighted sum of low-rank matrices. Experiments are run with the Prea API\footnote{\url{http://prea.gatech.edu/}}. We use a rank of 20, 30 anchor points which entails a global pseudo-rank of 600. Other hyperparameters are picked as recommended in \cite{Lee2013};
 \item I-Autorec \cite{Sedhain2015} trains one Autoencoder per item, sharing the weights between the different Autoencoders. We use 600 hidden neurons with the training hyperparameters recommended by the author.
\end{itemize}

In every scenario, we selected the highest possible rank which does not lead to overfitting despite a strong regularization. For instance, increasing the rank of BPMF does not significantly increase the final RMSE, idem for SVDFeature. Furthermore, we constrained the algorithms to run in less than two days. Similar benchmarks can be found in the litterature \cite{Li2016,Chen2015,Lee2013}.



\subsection{Data}
Experiments are conducted on MovieLens and Douban datasets.
The MovieLens-1M, MovieLens-10M and MovieLens-20M  datasets respectively
provide 1/10/20 millions discrete ratings
from 6/72/138 thousands users on 4/10/27 thousands movies. Side information for MovieLens-1M is the age, sex and gender of the user and the movie category (action, thriller etc.). Side information for MovieLens-10/20M is a matrix of tags $\mtx{T}$ where $T_{ij}$ is the occurrence of the $j^{th}$ tag for the $i^{th}$ movie and the movie category. No side information is provided for users.

The Douban dataset \cite{Hao2011} provides 17 million discrete ratings
from 129 thousands users on 58 thousands movies. Side information is the bi-directional user/friend relations for the user. The user/friend relation are treated like the matrix of tags from MovieLens. No side information is provided for items.

\paragraph{Pre/post-processing}
For each dataset, the full dataset is considered and the ratings are normalized from -1 to 1. We split the dataset into random 90\%-10\% train-test datasets and inputs are unbiased  before the training process: denoting $\mu$ the mean over the training set, $b_{u_i}$ the mean of the $i^{th}$ user and $b_{v_i}$ the mean of the $v^{th}$ item, U-CFN and V-CFN respectively learn from $r_{ij}^{unbiased} = r_{ij} -  b_{u_i}$ and $r_{ij}^{unbiased} = r_{ij} -  b_{v_i}$. 
%
The bias computed on the training set is added back while evaluating the learned matrix.

\paragraph{Side Information} In order to enforce the side information constraint, $Q \ll k_v \ll M$, Principal Component Analysis is performed on the matrix of tags. We keep the 50 greatest eigenvectors\footnote{The number of eigenvalues is arbitrary selected. We do not focus on optimizing the quality of this representation.} and normalize them by the square root of their respective eigenvalue: given $\mtx{T} = \mtx{P} \mtx{D} \mtx{Q}^T$ with $\mtx{D}$ the diagonal matrix of eigenvalues sorted in descending order, the movie tags are represented by $\mtx{Y} = \mtx{P}_{[1:M],[1:K']} \mtx{D}_{[1:K'],[1:K']}^{0.5}$ with $K'$ the number of kept eigenvectors. Binary representation such as the movie category is then concatenated to $\mtx{Y}$.

\begin{table*}[ht]
\begin{center}
\caption{RMSE with a training ratio of 90\%/10\%. The ++ suffix denotes algorithms using side information. When side information are missing, the N/A acronym is used. The * character indicates that the results were too low after four days of computation.}
\label{tab:RMSE}
\begin{tabular}{lcccc}
\hline
Algorithms  & MovieLens-1M                     & MovieLens-10M             & MovieLens-20M    & Douban\\
\hline
BPMF		&         0.8705  $\pm$ 4.3e-3  &          0.8213 $\pm$ 6.5e-4  &           0.8123 $\pm$ 3.5e-4     & 0.7133 $\pm$ 3.0e-4 \\
ALS-WR      &         0.8433  $\pm$ 1.8e-3  &          0.7830 $\pm$ 1.9e-4  &           0.7746 $\pm$ 2.7e-4     & 0.7010 $\pm$ 3.2e-4 \\
SVDFeature  &         0.8631  $\pm$ 2.5e-3  &          0.7907 $\pm$ 8.4e-4  &           0.7852 $\pm$ 5.4e-4     &              *      \\
LLORMA		&         0.8371  $\pm$ 2.4e-3  &          0.7949 $\pm$ 2.3e-4  &          0.7843 $\pm$ 3.2e-4      & 0.6968 $\pm$ 2.7e-4   \\
I-Autorec   &         \textbf{0.8305 $\pm$ 2.8e-3} &   0.7831 $\pm$ 2.4e-4  &          0.7742 $\pm$ 4.4e-4      & 0.6945 $\pm$ 3.1e-4 \\
\hline
U-CFN       &         0.8574  $\pm$ 2.4e-3  &          0.7954 $\pm$ 7.4e-4  &           0.7856 $\pm$ 1.4e-4     & 0.7049 $\pm$ 2.2e-4 \\
U-CFN++     &         0.8572  $\pm$ 1.6e-3  &                N/A            &                 N/A               & 0.7050 $\pm$ 1.2e-4 \\
V-CFN       &         0.8388  $\pm$ 2.5e-3  &          0.7767 $\pm$ 5.4e-4  &           0.7663 $\pm$ 2.9e-4     & \textbf{0.6911 $\pm$ 3.2e-4} \\
V-CFN++     &         0.8377  $\pm$ 1.8e-3  &  \textbf{0.7754 $\pm$ 6.3e-4} &   \textbf{0.7652 $\pm$ 2.3e-4}       &           N/A      \\
\hline
\end{tabular}
\end{center}
\end{table*}

\subsection{Error Function}


We measure the prediction accuracy by the mean of \emph{Root Mean Square Error} (RMSE).
Denoting $\mtx{R_{test}}$ the matrix test ratings and $\mtx{\widehat{R}}$ the full matrix returned by the learning algorithm, the RMSE is:
\begin{multline*}
RMSE(\mtx{\widehat{R}}, \mtx{R_{test}}) = \\
\sqrt{\frac{1}{|\mathcal{K}(R_{test})|}\sum_{(i,j)\in\mathcal{K}(R_{test})}(r_{test,ij}-\hat{r}_{ij})^2},
\end{multline*} 
where $|\mathcal{K}(R_{test})|$ is the number of ratings in the testing dataset.
Note that, in the case of Autoencoders, $\mtx{\widehat{R}}$ is computed by feeding the network with {\bf training} data. As such, $\hat{r}_{ij}$ stands for $nn(\mtx{u}_{train,i})_j$ for U-CFN, and  $nn(\mtx{v}_{train,j})_i$ for V-CFN.



\subsection{Training Settings}
We train 2-layers Autoencoders for MovieLens-1/10/20M and the Douban datasets. The layers have from $500$ to $700$ hidden neurons.  Weights are initialized  using the fan-in rule
\cite{LeCun1998}. 
Transfer functions are hyperbolic tangents. The Neural Network is optimized with stochastic backpropagation with minibatch of size 30 and 
a weight decay is added for regularization. 
Hyperparameters\footnote{Hyperparameters used for the experiments are provided with the source code.} are tuned by a genetic algorithm already used by \cite{Mary2007} in a different context. 

 \begin{table*}[ht]
    	\caption{RMSE computed by cluster of items sorted by their respective number of ratings on MovieLens-10M. For instance, the first cluster contains the 20\% of items with the lowest number of ratings. The last cluster far outweigh other clusters and hide more subtle results.}
    	        \label{tab:Side}
    \begin{subtable}{.5\linewidth}
      \centering
      \caption{MovieLens-10M (50\%/50\%)}
\begin{tabular}{lccc}
Interval \;& V-CFN & V-CFN++ & \%Improv. \\
\hline
0.0-0.2 &  1.0030   &  0.9938  & 0.96 \\
0.2-0.4 &  0.9188   &  0.9084  & 1.15 \\
0.4-0.6 &  0.8748   &  0.8669  & 0.91 \\
0.6-0.8 &  0.8473   &  0.8420  & 0.63 \\
0.8-1.0 &  0.7976   &  0.7964  & 0.15 \\
\hline
Full    &  0.8075   &  0.8055  & 0.25 \\
\hline
        \end{tabular}
                
    \end{subtable}%
    \begin{subtable}{.5\linewidth}
      \centering

                \caption{MovieLens-10M (90\%/10\%)}
\begin{tabular}{lccc}
Interval \;& V-CFN & V-CFN++ & \%Improv. \\
\hline
0.0-0.2 &  0.9539  &  0.9444  & 1.01 \\
0.2-0.4 &  0.8815  &  0.8730  & 0.96 \\
0.4-0.6 &  0.8487  &  0.8408  & 0.95 \\
0.6-0.8 &  0.8139  &  0.8110  & 0.35 \\
0.8-1.0 &  0.7674  &  0.7669  & 0.06 \\
\hline
Full    &  0.7767   &  0.7756  &  0.14 \\
\hline
	         \end{tabular}
    \end{subtable}
\end{table*}

\begin{table*}[ht]
	    	    \caption{Impact of the denoising loss in the training process. If we focus on the prediction (aka supervised setting), the autoencoder provides poor results. If we focus on the reconstruction with no masking noise (aka unsupervised setting), the Autoencoder already provides excellent results. By using a mixture of those techniques, the network converges to a better score.}	    	    	    
	    \label{tab:DAEimpact}
	    \begin{subtable}{.5\linewidth}
	      \centering
	        	        \caption{MovieLens-10M (90\%/10\%)}
	        \begin{tabular}{lccc|c}
	             & \; $\alpha$  \;& \;$\beta$  \; & \%Mask \;&\; RMSE \;\\
	\hline
	Supervised           &  0.91     & 0     & 0    & 0.8020\\
	Unsup.               &  0        & 0.54  & 0.25 & 0.7795\\
	Mixed                &  0.91     & 0.54  & 0.25 & 0.7768\\
	\hline
	        \end{tabular}

	    \end{subtable}%
	    \begin{subtable}{.5\linewidth}
	      \centering
	        	        \caption{MovieLens-20M (90\%/10\%)}
	        \begin{tabular}{lccc|c}
	             & \; $\alpha$  \;& \; $\beta$  \; & \%Mask \;&\; RMSE \;\\
	\hline
	Supervised           &  1        & 0     & 0.25 & 0.7982\\
	Unsup.               &  0        & 0.60  & 0    & 0.7690\\
	Mixed                &  1        & 0.60  & 0.25 & 0.7663\\
	\hline
	        \end{tabular}

	    \end{subtable} 

\end{table*}

\begin{figure*}[ht]
    \centering
    \includegraphics[width=0.75\textwidth]{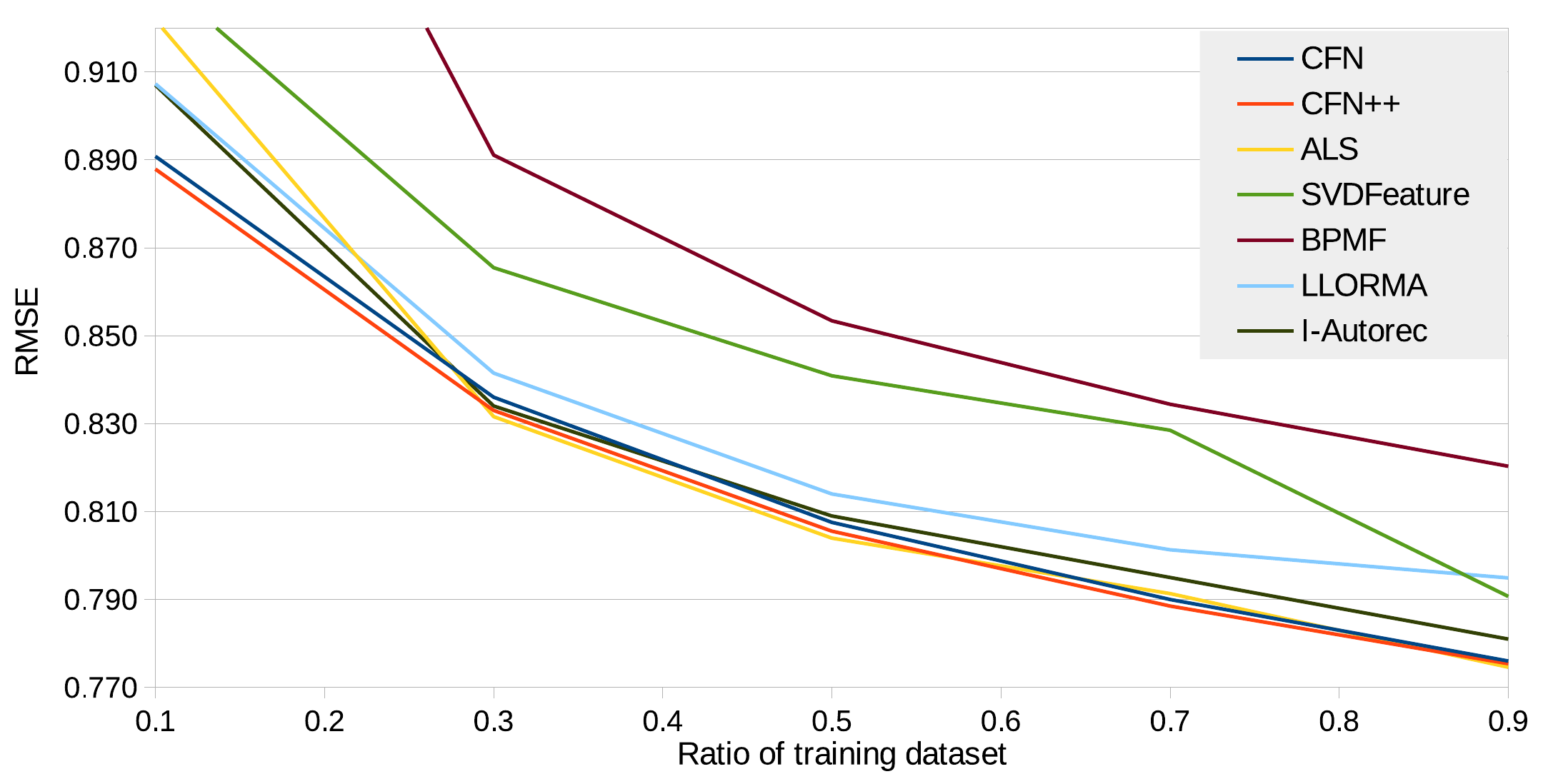}
    \caption{RMSE as a function of the training set ratio for MovieLens-10M. Training hyperparameters are kept constant across dataset. CFN and I-Autorec are very robust to a change in the density. On the other side, SVDFeature turns out to be unstable and should be fine-tuned for each ratio.}
    \label{fig:trainingRatio}
\end{figure*}

\subsection{Results} 

\emph{Comparison to state-of-the-art}.
Table \ref{tab:RMSE} displays the RMSE on MovieLens and
Douban datasets. Reported results are computed through $k$-fold cross-validation and
confidence intervals correspond to a 95\% range.
Except for the smallest dataset, V-CFNs leads to the best results; V-CFN is competitive compared to the state-of-the-art Collaborative Filtering approaches. To the best of our knowledge, the best result published regarding MovieLens-10M (training ratio of 90\%/10\% and no side information) are reported by \cite{Li2016} and \cite{Chen2015} with a final RMSE of respectively $0.7682$ and $0.7769$. However, those two methods require to recompute the full matrix for every new ratings. CFN has the key advantage to provide similar performance while being able to refine its prediction on the fly for new ratings. 
More generally, we are not aware of recent works that both manage to reach state of the art reslts while successfully integrated side information. For instance, \cite{Kim2014,Kumar2014} reported a global RMSE above $0.8$ on MovieLens-10M.

\begin{figure}[h]
    \centering
    \includegraphics[width=0.45\textwidth]{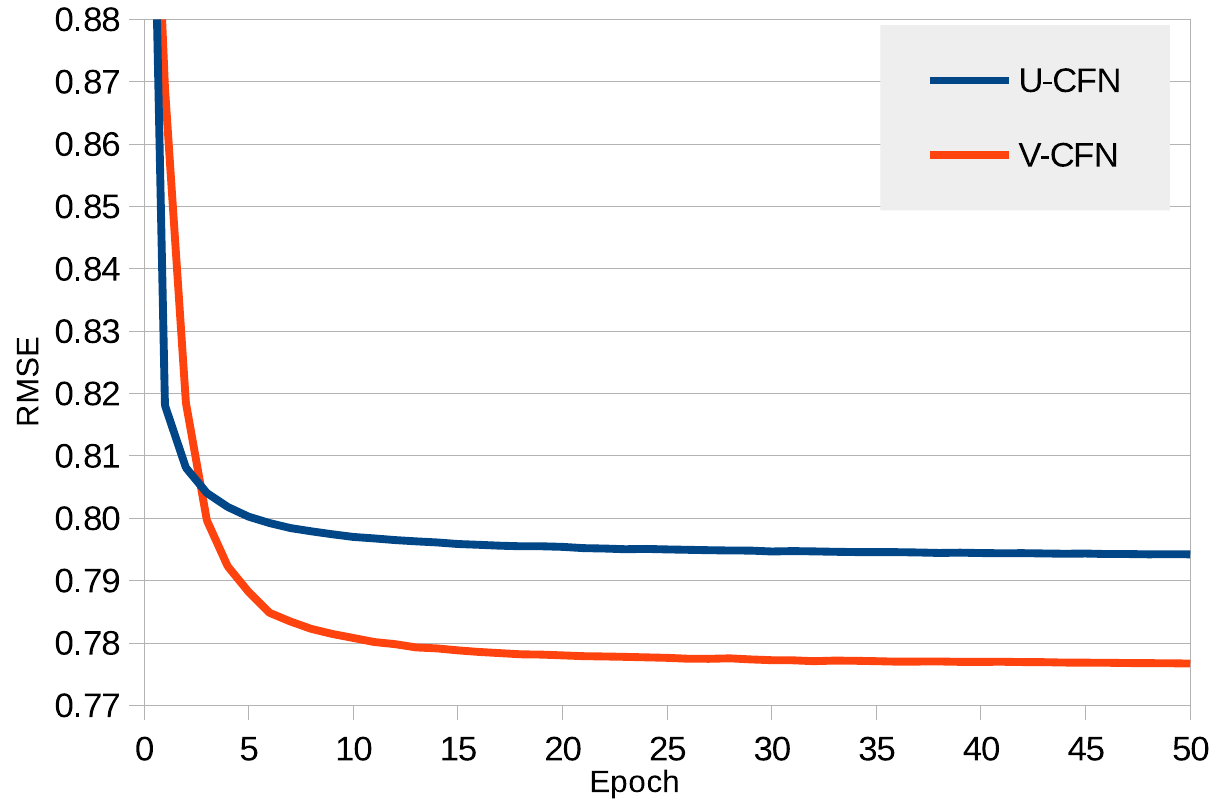}
    \caption{RMSE as a function of the numer of epoch for CFN for MovieLens-10M  (90\%/10\%). The network quickly converges to a very low RMSE and then refine its prediction upon epoches.}
    \label{fig:converge}
\end{figure}

%
%

Note that
V-CFN outperforms U-CFN. It suggests that the structure on the items is stronger than the one on users \textit{i.e.} it is easier to guess tastes based on movies you liked than to find some users similar to you. Of course, the behaviour could be different on some other datasets. The training evoluation is described in the Figure \ref{fig:converge}.

\emph{Impact of side information}.
At first sight at Table \ref{tab:RMSE}, the use of side information has a limited impact on the RMSE. This statement has to be mitigated: as the repartition of known entries in the dataset is not uniform, the estimates are biased towards users and items with a lot of ratings. For theses users and movies, the dataset already contains a lot of information, thus having some extra information will have a marginal effect. Users and items with few ratings should benefit more from some side information but the estimation bias hides them. 

In order to exhibit the utility of side information, we report in Table \ref{tab:Side}
the RMSE conditionally to the number of missing values for items.
As expected, the fewer number of ratings for an item, the
more important the side information.
This is very desirable for a real system: the effective use of side information to the new items is crucial 
to deal with the flow of new products.
A more careful analysis of the RMSE
improvement in this setting shows that the improvement is uniformly distributed
over the users whatever their number of ratings. This corresponds to the fact
that the available side information is only about items. To complete the picture, we train V-CFN on MovieLens-10M with either the movie genre
or the matrix of tags with a training ratio of 90\%/10\%. Both side information increase the global RMSE by 0.10\%
while concatenating them increases the final score by 0.14\%.
Therefore, V-CFN handles the heterogeneity of side information.

\emph{Impact of the loss}.
 The impact of the denoising loss is highlighted in Table \ref{tab:DAEimpact}: the bigger the dataset, the more usefull the de noising loss. On the other side, a network dealing with smaller dataset such as MovieLens-1M may suffer from masked entries. 

\emph{Impact of the non-linearity}.
We train I-CFN by removing the non-linearity to study its impact on the training. For fairness, we kept the $\alpha$, $\beta$, the masking ratio and the number of hidden neurons constant. Furthermore, we search for the best learning rates and L2 regularization throught the genetic algorithm. For movieLens-10M, we obtain a final RMSE of 0.8151 $\pm$ 1.4e-3 which is far worse than classic I-CFN. 

\emph{Impact of the training ratio}.
Last but not least, CFN remains very robust to a variation of data density as shown in Figure \ref{fig:trainingRatio}. It is all the more impressive that hyperparameters are first optimized for a training/testing ratio of 90\%/10\%. Cold-start and Warm-start scenario are also far more well-handled by Neural Networks than more classic CF algorithms. These are highly valuable properties in an industrial context.

\section{Remarks}
\label{sec:remarks}
\subsection{Source code}
Torch is a powerful framework written in Lua to quickly prototype Neural Networks. It is a widely used (Facebook, Deep Mind) industry standard. However, Torch lacks some important basic tools to deal with sparse inputs. Thus, we develop several new modules to deal with DAE loss, sparse DAE loss and sparse inputs on both CPU and GPU. They can easily be plugged into existing code. An out-of-the-box tutorial is  available to run the experiments. The code is freely available on Github\footnote{\url{https://github.com/fstrub95/Autoencoders_cf}} and Luarocks \footnote{luarocks install nnsparse}.

\subsection{Scalability}
One major problem that most Collaborative Filtering have to solve is scalability since dataset often have hundred of thousands users and items. An efficient algorithm must be trained in a reasonable amount of time and provide quick feedback during evaluation time. 

Recent advances in GPU computation managed to reduce the training time of Neural
Networks by several orders of magnitude. However, Collaborative Filtering deals
with sparse data and GPUs are designed to perform well on dense data.
\cite{Salakhutdinov2007,Sedhain2015} face this sparsity constraint by building
small dense Networks with shared weights. Yet, this approach may lead to
important synchronisation latencies. In our case, we tackle the issue by selectively densifying the inputs just before sending them to the GPUs cores without modification of the result of the computation. 
It  introduces an overhead on the computational complexity but this implementation
allows the GPUs to work at their full strength. In practice, vector operations overtake the extra cost. Such approach is an efficient strategy to handle sparse data which achieves a balance between memory footprint and computational time. We are able to train
Large Neural Networks within a few minutes as shown in Table
\ref{tab:Time}. For purpose of comparison, on MovieLens-20M with a 16 thread 2.7Ghz Core processor, ALS-WR (r=20) computes the final matrix within a half-hour with close results, SVDFeatures (r=64) requires a few hours, BPMF (r=10) and I-Autorec (r=600) require half a day, ALS-WR (r=200) a full day and LLORMA (r=20*30) needs several days with the Prea library. At the time of writing, alternative strategies to train networks with sparse inputs on GPUs are under development. Although, one may complain that CFN benefit from GPU, no other algorithm (except ALS-WR) can be easily parallelized on such device. we believe that algorithms that natively work on GPU are  auspicious in the light of the progress achieved on GPU.

\begin{table}[tbh]
\begin{center}
\caption{Training time and memory footprint for a 2-layers CFN without side information for MovieLens-10M (90\%/10\%). The GPU is a standard GTX 980. $Time$ is the average training duration (around 20-30 epochs). Parameters are the weight and bias matrices. Memory is retrieved by the GPU driver during the training. It includes the dataset, the model parameters and the training buffer. Although the memory footprint highly depends on the implementation, it provides a good order of magnitude. Adding side information would increase by around 5\% the final time and memory footprint.}
\label{tab:Time}	
\begin{tabular}{lcrrr}
\hline
Dataset  & CFN & \#Param & Time & Memory  \\
\hline
MLens-1M  & V &   8M &  2m03s  &   250MiB  \\
MLens-10M & V & 100M & 18m34s  & 1,532MiB  \\
MLens-20M & V & 194M & 34m45s  & 2,905MiB  \\
\hline
MLens-1M  & U &  5M  &  7m17s  &   262MiB  \\
MLens-10M & U & 15M  & 34m51s  &   543MiB  \\
MLens-20M & U & 38M  & 59m35s  & 1,044Mib  \\
\hline	
\end{tabular}
\end{center}
\end{table}

\subsection{Future Works}
\label{sec:ccl}

Implicit feedback may greatly enhance the quality of Collaborative Filtering
algorithms \cite{Koren2009,Rendle2010}. For instance, Implicit feedback would be incorporated to CFN by feeding the Network with an additional binary
input. By doing so, \cite{Salakhutdinov2007} enhance the quality of prediction for Restricted Boltzmann Machine on the Netflix Dataset.
Additionally, Content-Based Techniques with Deep learning such as \cite{Van2013,Wang2014b} would be plugged to CFN.
The idea is to train a joint Network that would directly link the raw item features to the ratings such as music, pictures or word representations.
As a different topic, V-CFN and U-CFN sometimes report different errors.
This is more likely to happen when
they are fed with side information. One interesting work would be to combine a
suitable Network that mix both of them.
Finally, other metrics exist to estimate the quality of Collaborative Filtering to fit other real-world constraints. 
Normalized Discounted Cumulative Gain \cite{Jarvelin2002} or F-score are sometimes preferred to RMSE and should be benchmarked.

\section{Conclusion} 
In this paper, we have introduced a Neural Network
architecture, aka CFN, to perform Collaborative Filtering with side
information. Contrary to other attempts with Neural Networks, this joint Network integrates side information and learns a non-linear 
representation of users or items into a unique Neural Network.
This approach manages to beats state of the art results in CF on both MovieLens and Douban datasets. It performs excellent results in both cold-start and warm-start scenario. CFN has also valuable assets for industry, it is scalable, robust and it successfully deals with large dataset.  
Finally, a reusable source code is provided in Torch and hyperparameters are provided to reproduce the results.

 \subsection*{Acknowledgements}
 
 The authors would like to acknowledge the stimulating environment provided by SequeL research group, Inria and CRIStAL. This work was supported by French Ministry of Higher Education and Research, by CPER Nord-Pas de Calais/FEDER DATA Advanced data science and technologies 2015-2020, the Projet CHIST-ERA IGLU and by FUI Herm\`{e}s. Experiments were carried out using Grid'5000 tested, supported by Inria, CNRS, RENATER and several universities as well as other organizations. 


\bibliographystyle{IEEEtran}
\bibliography{biblio}

\section{Appendix}

\subsection{Genetic Algorithm}
We use the following genetic algorithm \cite{Mary2007} to find the hyperparameters of our model.
The cross-over of two individuals $\mtx{x}$ and $\mtx{y}$ gives birth to two new individuals  $\frac{2}{3}\mtx{x} + \frac{1}{3}\mtx{y}$ and $\frac{1}{3}\mtx{x} + \frac{2}{3}\mtx{y}$.
The mutation of one individual is obtained by using an isotropic Gaussian law with the mean centred on the current values of parameters and a standard deviation of $\frac{\sigma}{\sqrt[•]{•}sqrt[n]{d}}$ with $n$ the number of individuals and $d$ the dimension of the space.
Let $\lambda_1$, $\lambda_2$, $\lambda_3$ and $\lambda_4$ be such that $\lambda_1 + \lambda_2 + \lambda_3 + \lambda_4 = 1$.
Once an initial population of $n$ individuals is created, we proceed as follow at each iteration:
\begin{itemize}
\item We copy $n\lambda_1$ best individuals (Set $S_1$)
\item We apply the cross-over rule to the $\frac{n}{2}\lambda_2$ following best individuals with randomly picked individuals in $S_1$ 
\item We mutate $n\lambda_3$ randomly picked individuals in $S_1$
\item We generate $n\lambda_4$ new individuals
\end{itemize}

\begin{table}[H]
\begin{center}
\caption{Gene description for CFN. Hyperparameters of the genetic algorithms were $n=20$, $\sigma = 0.08$, $\lambda_1 = 1/10$, $\lambda_2= 2/10$, $\lambda_3= 3/10$, $\lambda_4= 4/10$}
\vskip 0.1in
\begin{tabular}{lc}
\hline
CFN hyperparameters  & Probabilistic law-1M    \\                
\hline
$alpha$             &  U[0.8,1.2]\\
$beta$              &  U[0,1] \\
masking ratio       &  U[0,1] \\
bottleneck size     &  U[500,700]\\
learning rate       &  U[0,0.5]\\
learning rate decay &  U[0,0.5] \\
weight decay        &  U[0,0.5] \\
\hline
\end{tabular}
\end{center}
\vskip -0.1in
\label{tab:gene}
\end{table}

\end{document}